# Study of ZrO$_2$ Thin Films Deposited at Glancing Angle by Magnetron Sputtering under Varying Substrate Rotation

## By

R. B. Tokas[1*], S. Jena[1], J. S. Mishal[2], K. Divakar Rao[2], S. R. Polaki[3], C. Pratap[2], D. V. Udupa[1], S. Thakur[1], Sanjiv Kumar[4] and N. K. Sahoo[1]

[1]Atomic & Molecular Physics Division, Bhabha Atomic Research Centre, Trombay, Mumbai 400085, India

[2]Photonics & Nanotechnology Section, Atomic & Molecular Physics Division, Bhabha Atomic Research Centre Facility, Visakhapatnam 530012, India

[3]Indira Gandhi Centre for Atomic Research, Kalpakkam-603102, India

[4]National Centre for Compositional Characterization of Materials, BARC, Hyderabad 500062, India

*Email: tokasstar@gmail.com

Address for Correspondence:

R. B. Tokas

Atomic & Molecular Physics Division
Bhabha Atomic Research Centre
Trombay, Mumbai 400 085, India
Phone: 91-22-25590341
E-mail: tokasstar@gmail.com

# Study of $ZrO_2$ Thin Films Deposited at Glancing Angle by Magnetron Sputtering under Varying Substrate Rotation

**Abstract:** Glancing angle deposited oxide thin films are becoming popular to fabricate challenging interference devices, sensors, thermal barrier coatings etc. Such films depict interesting tunable optical, morphological, mechanical and structural properties due to their tunable porous columnar micro-structure. By varying substrate rotation and angle of deposition, columnar microstructure can be varied which leads to the varying properties of such films greatly. In the present work, a set of $ZrO_2$ thin films have been deposited at 82º glancing angle of deposition at several substrate rotation speeds and at 0º angle (normal deposition). The effect of substrate rotation on optical, structural, morphological properties and residual stress has been studies thoroughly. Refractive index estimated from ellipsometric measurement and suitable modeling depicts an interesting decreasing behavior with substrate rotation and has been explained in the light of varying columnar structure with substrate rotation. Refractive index of GLAD $ZrO_2$ films varies between and 1.901 to 2.011. Normally deposited film exhibits refractive index value of 2.178 which is substantially greater than that of GLAD films. Lowering in refractive index of glancing angle deposited films is the attribute of dominant atomic shadowing at glancing angles. Further, correlation length which is the representative of surface grain size was obtained from suitable modeling of atomic force microscopy data and it exhibits a decreasing trend with substrate rotation. The trend has also been attributed to the varying columnar microstructure with substrate rotation. All the glancing angle deposited $ZrO_2$ films possess root mean square roughness between 4.6 and 5.1 nm whereas normally deposited film depicts 1.0 nm rms roughness. Dominant atomic shadowing is responsible for high roughness of

GLAD films. Both glancing angle and normally deposited films exhibit preferential growth of monoclinic phase oriented in different directions. GLAD films also depict a tetragonal peak which has been attributed to the fine nano-crystallite size (~13 nm). Residual stress depicts a great switching from large compressive to small tensile as the deposition angle switches from normal to glancing angle and GLAD films except deposited at highest substrate rotation (4 rpm) depict an increasing tensile stress with substrate rotation. Lowering in stress for GLAD films and its trend with substrate rotation has been explained in terms of varying inter-molecular forces with inter-columnar distance.

## 1. Introduction

Glancing angle deposition (GLAD) of thin films has been an attractive field for researchers and technologists for recent years due to their fascinating physical and chemical properties. In this technique[1-3], substrate is kept at an angle with respect to incoming ad-atoms flux and by employing the variation of substrate rotation, angle of deposition and rate of deposition, different microstructures depicting unique physical and chemical properties could be achieved. The physical properties i.e., refractive index, thermal conductivity, electrical conductivity, wetting properties etc., could be engineered using such GLAD coatings. In this technique, there occurs a columnar micro-structure with high porosity which is the result of dominant ballistic atomic shadowing with limited ad-atom diffusion on substrate surface. This porous microstructure is responsible for unique properties, which are exploited for various applications i.e., energy harvesting and storage devices, water splitting, fuel cell and hydrogen storage, sensors, optical applications, wetting and micro fluidics, biomaterial and biosensors[2-4]. Porosity in thin films due to atomic shadowing is very high near glancing angle and varies non-linearly with angle of deposition[5]. GLAD could be carried out by electron beam (EB) and thermal evaporation, pulsed laser deposition, plasma enhanced chemical vapor deposition and magnetron sputtering techniques[3]. Out of these technique, EB and thermal evaporation are known to produce the most porous nano rod structure due to the directional movement and low scattering of ad-atoms. Other techniques, especially magnetron sputtering being an energetic process produce merged porous microstructure rather than separate nano rods for glancing angles ~ 80º. Columnar morphology in GLAD films is influenced by substrate rotation, rate of deposition, deposition technique etc. Among all substrate rotation and deposition method affect the columnar morphology greatly. Therefore understanding of properties of GLAD films fabricated by different method is of great importance.

$ZrO_2$ alone or with stabilizing additives i.e., $Y_2O_3$, CaO etc., depicts excellent optical, mechanical, electrical, chemical and thermal properties which enable it to become an important material for applications in various fields. [6] From optical applications point of view, $ZrO_2$ has high refractive index, high pulse laser induced damage threshold and broad low absorption transparency region from near UV to mid IR wavelength region[7-9]. Glancing angle deposition of $ZrO_2$ thin films fabricated by different deposition techniques under varying deposition parameters, have also been explored by many researcher to characterize their various properties. Sumei Wang, et. al., had investigated the influence of angle of deposition on optical and columnar micro-structural properties and birefringence in refractive index of $ZrO_2$ thin films deposited by EB evaporation[7]. Levichkova, et al., have discussed the influence of angle of deposition on DC electrical conductivity, micro-hardness, optical anisotropy in EB evaporated GLAD $ZrO_2$ thin films[10]. Yong Jun Park, et al., have realized a wideband circular polarization reflector using GLAD $ZrO_2$ and other oxide thin films by EB evaporation[11]. I. J. Hdgkinson, et al., have demonstrated and measured biaxial anisotropy in refractive index of GLAD $ZrO_2$ thin films at stationary substrate by EB evaporation[12]. Most of the work done on GLAD $ZrO_2$ thin films has been carried out using EB evaporation. As mentioned above, Sputtering depicts different micro-structures than EB evaporation in GLAD configuration; especially less porosity at a given angle of deposition, so it is of great importance of characterize GLAD $ZrO_2$ thin films deposited by magnetron sputtering under varying deposition parameters. In present work, we report an extensive study of optical, morphological and micro-structural properties and residual stresses of GLAD $ZrO_2$ thin films deposited by RF magnetron sputtering technique with varying substrate rotation. Spectroscopic ellipsometry has been used for optical characterization. Atomic force microscopy (AFM) and field emission scanning electron microscopy (FESEM) has been used for surface and cross-section morphological characterization. Grazing incidence X-ray ray diffraction (GIXRD) has been employed for identifying crystalline phases and orientations. Stresses have been estimated using Fizeau interferometer. All the parameters obtained from different characterization of GLAD films have been compared with that of normally deposited (ND) $ZrO_2$ thin film.

## 2. Experimental Details

$ZrO_2$ thin films have been deposited by RF magnetron sputtering in normal and GLAD configuration. GLAD films have been deposited at an angle of deposition of 82° on Si (111) substrate with substrate rotation speed 0.5, 1.0, 1.5, 3.0 and 4.0 revolutions per minute (rpm) and have been designated as Samp-1, Samp-2, Samp-3, Samp-4 and Samp-5 respectively. ND $ZrO_2$ film has been designated as Samp-6. Actual photograph of GLAD sputtering set-up which utilizes a DC stepper motor for substrate rotation is shown in Fig. 1. Substrate to target distance was optimized to get a stable plasma and good deposition rate and kept 70 mm. In order to reduce the contaminations in coatings, prior to deposition, a substantially low pressure of ~ $1.5 \times 10^{-5}$ mbar inside the sputtering chamber was achieved using turbo-molecular vacuum pump baked by rotary pump. Deposition pressure was maintained at ~ $2.8 \times 10^{-3}$ mbar with a flow rate of 19 SCCM of sputtering gas (Ar) inside the vacuum chamber and controlled by mass flow controller. Sputtering was carried out at RF power of 250 Watts to give a deposition rate ~ 2 Å/s for ND film. MDC make quartz crystal controller was used to monitor and control the rate of deposition. $ZrO_2$ target was pre-sputtered for 300 seconds to remove any possible contamination from the surface.

Optical characterization of GLAD $ZrO_2$ thin films has been carried out by a SEMILAB make rotating polarizer type spectroscopic ellipsometer (GES5E). Ellipsometric measurements were performed in the wavelength range of 200-800 nm at 70° angle of incidence on thin films. Morphological measurements of the films were carried out by NT-MDT, Russia make Solver P47H Atomic Force Microscopy (AFM) system in tapping mode with a super sharp DLC tip grown on silicon cantilever having radius of curvature 1-3 nm; fundamental resonance frequency 193 kHz and spring constant of 8.1 N/m. In order to explore microstructure and determine thickness of films, cross-sectional morphology was recorded by field emission scanning electron microscopy (FESEM), Zeiss Supra 55VP system. Stoichiometry of the films was determined by using Rutherford backscattered measurement using 2.0Mev proton beam. Crystal structure of GLAD $ZrO_2$ thin films was examined by recording GIXRD pattern by Rikaku Ultima-IV XRD machine using Cu Kα X-ray radiation ( $\lambda=1.5402$ Å). GIXRD patterns were recorded in 2θ geometry with a scan speed of 1°/minute in angular range of 20°-50°. The grazing angle of X-ray on the thin film samples was kept fixed ~ 0.5°. Curvature of Si substrates was measured before and after deposition of thin films using laser Fizeau interferometer. Finally Stoney formulation was used to evaluate residual stress in such thin films.

## 3. Results and discussion

### 3.1. Optical Properties

Optical properties of $ZrO_2$ thin films have been estimated by ellipsometry technique. In ellipsometry, change in amplitude and phase difference between parallel (p) and perpendicular (s) components of electric field of light reflected from surface of the test sample are measured. When plane polarized light is incident on the surface of sample near Brewster angle, it gets reflected in 'p' and 's' components with different amount of amplitude and phase. The ratio of reflection coefficient of 'p' and 's' components of light and phase difference between them are related as follows[13,14]:

$$\rho = \frac{r_p}{r_s} = \tan(\psi)\exp(i\Delta) \tag{1}$$

Where $\psi$ and $\Delta$ stand for amplitude ratio and relative phase change between 'p' and 's' components of electric field amplitude of light respectively. In rotating polarizer type ellipsometry, the time dependent signal, $I(t)$ at each wavelength is described by the general formula[15]:

$$I(t) = I_0(1 + \alpha\cos(2wt) + \beta\sin(2wt)) \tag{2}$$

Where $\omega$ is the rotational frequency of the polarizer and $\alpha$ and $\beta$ are the normalized Fourier coefficients with which the ellipsometric parameters $\psi$ and $\Delta$ are related through following expression:

$$\alpha = \frac{\tan^2\psi - \tan^2 A}{\tan^2\psi + \tan^2 A} \quad \text{and} \quad \beta = \frac{2\tan\psi\cos\Delta\tan A}{\tan^2\psi + \tan^2 A} \tag{3}$$

where A is the angle of analyzer. Even though $\psi$ and $\Delta$ are physically more meaningful ellipsometric parameters, the more directly obtained ellipsometric parameters $\alpha$ and $\beta$ (at analyser angle of $45°$) have been used for present analysis.

It is a known fact that thin films deposited by physical vapor deposition comprise voids (air) and the fraction of voids in the films depends on deposition technique and process parameters. Magnetron sputtering is an energetic deposition technique and the films grown by it are dense having low fraction of voids. However, in present study, $ZrO_2$ films were grown in GLAD configuration and void fraction is significantly high due to atomic shadowing effects of ad-atoms. Considering the porous nature of such films, each film has been modeled as two layers of $ZrO_2$ consisting different void and material component fractions and following the same dispersion relation for refractive index. Top layer (surface) is more porous due to being in direct contact with air. The model thickness of top layer ranges from 17-22 nm. Underlying layer is modeled as bulk layer with a homogeneous mixture of $ZrO_2$ and void with variable fraction of material and void component for different samples. In Fig. 2, model layer structure for representatives of $ZrO_2$ thin films, Samp-1 and Samp-6 are shown in inset. Since sputtering is an energetic deposition technique, the porosity introduced in thin films is not large even in GLAD configuration[14] in bulk layer in contrast to EB deposited films at the same angle. Hence, such films have been assumed optically isotropic for ellipsometric modeling. The calculation of effective refractive index has been done using Bruggman effective medium approximation[16]. Cody-Lorentz model[17] was used as dispersion relation of $ZrO_2$ material component in all the layer structure. Cody-Lorentz dispersion model is Kramer-Kroning consistent and is an improvement of Tauc-Lorentz dispersion formulation[18] and can be used to model amorphous and polycrystalline materials. The main differences between the two models are a modified joint density of state and the fact that Code-Lorentz relation has weaker exponential absorption below band gap. At, E0, Γ, Eg, Ep, Eu, Et and ε(∞), are eight model parameters of Cody-Lorentz rule to describe the dispersive complex dielectric constant/refractive index of material. Here At is the transition strength, E0 is the peak transition energy, Γ is broadening energy, Eg is the band gap, Ep is a transition energy that separates the absorption onset behavior from the Lorentz oscillator behavior, Eu is the slope of Urbach tail that represents disorder in the material, Et is the demarcation energy between the Urbach tail transitions and band to band transitions and ε (∞) is dielectric constant at infinite energy[18].

Taking initial trial values of layer thicknesses, fraction of voids and components and dispersion model fitting parameters, measured ellipsometric spectra were fitted by minimizing root mean squared value between measured and theoretically generated ellipsometric parameters α (45º) and β (45º) as follows:

$$RMSE = \left[\{1/(2N-P)\}\sum_{i}^{N}\left\{(\alpha_i^{\exp}(45°) - \alpha_i^{cal}(45°))^2 + (\beta_i^{\exp}(45°) - \beta_i^{cal}(45°))^2\right\}\right]^{1/2} \quad (4)$$

Where *N* is the number of data points and P is the number of fitting parameters. Apart from dispersion relation parameters, film thickness was also taken as a fitting parameter. In Fig. 2 (a) and 2(b), experimental and fitted curves of α (45º) and β (45º) are shown. It can be noted that the theoretical curves fit very well with measured curves. It justifies the selection of Coy-Lorentz dispersion relation and correctness of chosen layer structures. All the films except Samp-6 (ND), have been modeled using 2-layers. Samp-6 depicts a dense micro-structure with a low void fraction of 1.4% and ellipsometric curves fits very well with single layer structure only. Thickness, void fraction, refractive index of each layer, effective refractive index and extinction coefficient derived from ellipsometric analysis are listed in Table-1 for all the films. Dispersive refractive indices of all the $ZrO_2$ films are plotted in Fig. 3(a). Samp-6 depicts greater refractive index (2.178) compared to the GLAD films (1.901-2.011). The reduction in refractive index of GLAD films is the attribute of prominent ballistic atomic shadowing effects which introduces porosity in the films. As can be seen from Table-1, the least difference between GLAD and ND film is 0.167 which is smaller compared to GLAD films deposited by EB evaporation under similar deposition angle configuration[19]. As mentioned in section-1, ad-atoms in sputtering possess much higher average energy compared to EB deposition and high ad-atoms energy gives them mobility to hope over the surface of the substrate. Moreover deposition pressure in sputtering is relatively high (~$10^{-3}$ mbar) which leads to the scattering of ad-atoms by sputtering gases. Consequently, such scattered ad-atoms become randomly directed up to some extent and hence can fill some of the voids generated due to shadowing effects and results less porosity in sputtered GLAD films compared to EB deposited GLAD film under the same deposition angle. Effective rate of deposition was calculated by measuring post deposition film thickness by ellipsometry and deposition time. As listed in Table-1, rate of deposition reduces drastically for GLAD films compared to ND film. The rate of deposition for ND film is 2.06 Å/s whereas for GLAD films, it lies between 0.64 and 0.74 Å/s. Such reduction for GLAD films is also supported by earlier results published by many researchers[2,3] and is the attribute of reduction in sticking coefficient due to high incident angle of ad-atoms on substrate[20]. Deposition rate increases monotonically with the increase in substrate rotation (0.64-0.74 Å/s). Such results suggest that column length to diameter ratio increases with substrate rotation. So it can be inferred that with the increase in substrate rotation, column diameter decreases and length increases. Fig. 3(b) presents the variation of refractive index of GLAD $ZrO_2$ films with substrate rotation and it decreases with increase in rotation speed. Initially, refractive index decreases with higher slope and for rotation speed $\geq$ 3 rpm, decreasing rate of index decreases. There are earlier reports on effect of rotation speed on columnar microstructure but most of them are for EB evaporation[21]. In their case, there was a transition from helical columns to strait pillars as the speed exceeds 0.1222 rpm and structure of columns is fibrous. In our case, as the speed increases, the columns tend to become more cylindrical and continuous solid pillars. The decrease in refractive index with substrate rotation is corroborated by the trend of deposition rate for GLAD films. Although for higher rotation speeds, refraction index varies slowly and it suggests the saturation of evolution of columnar microstructure with the increase in substrate rotation. It can also be noted from Table-1 that extinction coefficient (k) of GLAD $ZrO_2$ films ($\sim 10^{-3}$) is one order higher than that of ND film ($\sim 10^{-4}$). Higher k values for GLAD films could be the results of higher scattering losses in UV region due to high roughness which will be discussed in next section.

*3.2 Surface and cross-sectional Morphological properties*

Surface morphology of $ZrO_2$ films was investigated by AFM using DLC tip grown on Si silicon cantilever. Having high hardness, DLC tip offers high resistance to abrasion and changes in its geometry due to measurements. So, it adds an accuracy and consistency to the measurement and results. For AFM measurements, tapping mode which utilizes a very sensitive resonance based detection technique has been used. In this mode, AFM tip touch the sample once only in one oscillation and artifacts due to tip-sample contact get (creeping, scratches in sample, abrasion to tip) reduced[22].

For quantitative analysis of surfaces of such films, one dimensional height-height correlation function which is a second order surface statistical parameter, has been computed from AFM measurements using following formulations[20,23]

$$H(r_x) = \frac{1}{N(N-m)} \sum_{l=1}^{N} \sum_{n=1}^{N-m} (z_{n+m,l} - z_{n,l})^2 \quad (5)$$

Here m = $r_x/\Delta x$. The height-height correlation function (HHCF) thus can be evaluated in a discrete set of $r_x$ value separated by sampling interval $\Delta x$. In above formulation, scan data points have been taken equal in x and y direction (fast and slow scan direction). HHCF prominently describes the grain structure in thin films. To get the useful surface parameters, we have fitted the measured function with calculated one for self affine surfaces using following Gaussian function[24]

$$H(r) = 2\sigma^2 \left[1 - \exp(-(r/\xi)^{2\alpha})\right] \quad (6)$$

Where σ is rms roughness or often termed as interfacial width and describes the fluctuations of heights around the mean plane. Exponent α is "Herst" or roughness exponent and accounts for short range roughness. It describes the grain morphology and depicts the wiggliness of the local slope on the surface. Higher value of α (≤ 1) account for less high frequency contribution to line edge roughness[25] or we can call it smoother line edge. ξ is lateral correlation length and it describes the largest distance in which the height is still correlated on the surface. So ξ denotes the distance after which the edge points are considered uncorrelated. In Fig. 4 (a), measured HHCFs are shown for all the ZrO$_2$ films. Higher counts of HHCF stand for higher surface rms roughness. It can be noted that for all GLAD films, counts of HHCF are close to each other whereas for ND film, HHCF value is relatively lower. In Fig. 4(b), measured and fitted HHCF curves for a representative of ZrO$_2$ films, Samp-1 are shown. The parameters derived from the fitting of HHCF are listed in Table-2. RMS roughness (σ) for GLAD ZrO$_2$ films varies between 4.4 and 5.1 nm and there is no significant and systematic influence of substrate rotation speed on RMS roughness. RMS roughness for normally deposited film is 1 nm which is 4 to 5 times lower than that of GLAD films. It is reported that re-emission and atomic shadowing[20,26] are the two main factors contributing to the roughness of thin film surfaces. In present case, for ND film, sticking coefficient of ad-atoms is highest which attributes to lower re-emission of ad-atoms and hence lower smoothing effect. Roughening effects due to atomic shadowing are negligible for ND film. Collectively, both the effects do not contribute to the roughening of ND efficiently. On the other hand, atomic shadowing for GLAD films is high and gives substantially high roughening effects to the films. However, smoothing effect increases due to increase in re-emission caused by decrease in sticking coefficient, the collective effect give a great rise to the roughening of surfaces of GLAD films. A little variation in roughness may be because of variation of rms roughness of different substrates. Variation of correlation length ($\sigma$) which is a measure of surface grain size is shown in Fig. 5. $\sigma$ decreases gradually with substrate rotation and hence it can be said that average grain size decrease with substrate rotation. Decrease in thin films. Decreasing trend of column diameter is already explained in section 3.1. As mentioned in section 3.1, the statement, "the columns tend to become more cylindrical and continuous solid pillars with rotation speed" is corroborated by AFM images shown in Fig. 6. For lower rotation speeds (0.5 and 1.0 rpm), grains have irregular and non circular shape (Fig. 6(a, b). 3-D zoomed images in Fig. 6 shown on right side also depicts non-circular grains. As the rotation speed increases, grain tends to become circular as shown in Fig. 5(c, d, e). It can also be noted from AFM surface morphologies (Fig. 6) that void fraction tends to increase with increase in substrate rotation. Increasing trend of void fraction on surface corroborates the variation of refractive index (decreasing trend) as discussed in section 3.1. The values of roughness exponent α are listed in Table-2 and it remains almost unchanged with substrate rotation except Samp-1 (0.5 rpm). Value of α for samp-1 is 0.86 and the lowest among all the GLAD films. It signifies the relatively more wiggly surface i.e. slope on the edges of grains is sharp or high. Fig. 6 also depicts a wiggly surface for Samp-1 (0.5 rpm) and Samp-6 (ND). AFM image of GLAD films other than Samp-1 depict a smooth variation at grain boundaries. Higher grain density for Samp-1 and 6 (ND) is responsible for higher wiggly surface and hence lower α value.

FESEM cross-sectional morphology of all the $ZrO_2$ films is shown Fig. 7. It can be seen that All the GLAD films depict a porous erect columnar structure which is caused by ballistic atomic shadowing during the growth of the films. However columns are not distinct and are merged to each other. The reason of the same is already explained in sec. 3.1. Thickness of the films has also been measured from cross-sectional morphological images and is listed in Table-2. It can be noted that thickness values obtained from FESEM measurements are consistent with the values obtained from ellipsometric analysis. It also justifies the methodology used for analysis of ellipsometric measurements.

*3.3 Study of structural properties*

Since $ZrO_2$ thin films were deposited without supplying additional oxygen, stoichiometry of the films was of concern. In Fig. 8, measured and simulated backscattered spectra of proton are shown for Samp-6 (ND). From suitable modeling of back scattered spectra, formation of fully stoichiometric (oxidic) $ZrO_2$ films was confirmed. The backscattered measurements also depict a small peak of Hf having a concentration < 1 %. Zr and Hf are very similar elements and it is very difficult to separate them. In our oxide targets, impurity % of $HfO_2$ is <1%. X-ray diffraction patterns taken in grazing incidence configuration are shown in Fig. 9 for all the films. Major X-ray diffraction Peaks for GLAD $ZrO_2$ films were observed at $2\theta = 24°, 28.1°, 30.2°, 31.5°, 34°, 34.4, 44.8$ and assigned to the crystalline planes of monoclinic and tetragonal phase with directions m(110), m(-111), t(101), m(111), m(002), m(020), m(112) respectively and plane direction are in agreement with 83-0936 and 81-1545 standards of **JCPDS.[]**. As shown in Fig. 9, GLAD films depict a preferential growth for crystalline plane of monoclinic phase in m (110) direction. Generally for normal deposition of $ZrO_2$ films, the most intense XRD peak corresponding to m (-111) direction around~ 28.2 is observed but in GLAD films that peak is almost absent having only the signature. Since different crystal direction or face grows at different rates, the vertical growth (perpendicular to substrate) rate is sensitive to the orientation of the crystal. The crystallites having the orientation with the greatest vertical growth are most likely and a preferential growth in that direction of crystallographic planes occurs. In GLAD films, the ad-atom strikes substrate surface at glancing angle and leads to the preferential growth in m (110) direction. Another thing to be noticed is the presence of tetragonal phase positioned at 30.2° with directions (101) and it is not common to get tetragonal phase of $ZrO_2$ at room temperature. Tetragonal peaks for all GLAD films were fitted with a Lorentzian and FWHM was found ~ 0.63±0.01 in degree. Crystallite size was calculated using Scherrer equation[27] ( $t = 0.94\lambda/(B_{1/2}\cos\theta)$ and found (13.6 ± 0.2) nm. It clearly indicates that the tetragonal $ZrO_2$ is in nanocrystalline form. At room temperature monoclinic is the most stable phase of $ZrO_2$ due to thermo dynamical conditions[28] and tetragonal is high temperature phase which generally occurs at temperature ≥ 600°. Researchers had reported that high temperature tetragonal phase of $ZrO_2$ could also be obtained at room temperature without adding any stabilizing doping by two methods; high pressures[29] and fabricating material with very fine crystal grains. It is reported[30,31] that a small crystallite size could reduce tetragonal to monoclinic phase transformation temperature substantially (room temperature). The critical diameter of tetragonal phase crystallite to stabilize the tetragonal at room temperature is ≤ 30 nm. Since in present case, crystallite size is ~13.6 nm, the presence of tetragonal phase could be attributed to fine tetragonal crystallites. The crystallite size effect could be explained in terms of the lower surface energy of the tetragonal phase (770 erg/cm$^2$) compared to that of rather stable monoclinic phase (1130 erg/cm$^2$), which could compensate for the chemical free energy. For dominant monoclinic peak at 24º, FWHM of the peak was found ~ 0.45 degree and crystallite size was in the range 18.5 to 19.5 nm which is more than that for tetragonal phase grains. For ND films film (Samp-6), XRD peaks were observed at 2θ = 27.6º, 33.7 º, 35.2 º and 44.6 º corresponding to the monoclinic phase with directions m (-111), m (002), m (200), and m (112) respectively. The peaks are shifted to lower 2θ value and the shift may be attributed to high compressive residual stress due to dense microstructure. ND film depicts preferential growth in m (-111) directions. For such film, energetic ad-atoms strike on substrate vertically and cause the greatest vertical growth of m (-111) reflection planes. It can be noted that the peak at m (-111) is asymmetric and extended to higher 2θ values. This is due to the merging of small peaks of t (101) and m (111) with intense and broad peak m (-111). So the films deposited normally and at glancing angle show different preferential crystal growth directions with the presence of tetragonal phase.

*3.4. Residual stress analysis*

Residual stress in ZrO$_2$ thin films has been derived from famous Stoney's equation[32] as follows

$$\sigma = \frac{1}{6}\left(\frac{E_s}{1-\nu_s}\right)\left(\frac{1}{R_2} - \frac{1}{R_1}\right)\left(\frac{t_f^2}{t_s}\right) \qquad (7)$$

Where $R_1$ and $R_2$ are radii of the substrate curvature before and after film deposition respectively; $E_s$ is the Young's modulus of elasticity; $\nu_s$ is the Poisson's ratio of the substrate; $t_s$ and $t_f$ are thicknesses of the substrate and film respectively. For the validity of Stoney's formula, thickness of the film should be much less than that of substrate ($t_f \ll t_s$). For present analysis sign convention for compressive and tensile stress has been taken negative and positive respectively. Surface profiles were obtained by analysis of two beam interference fringes formed between the substrate and an optical flat standard in a laser Fizeau interferometer[33]. Surface curvature has been estimated from surface spherical shape components which have been determined by fitting the surface profile obtained from interferometer with Zernike polynomial function[33]. Mean radius of curvature has been calculated from curvature profiles using following equation[34]:

$$R = \frac{r^2}{2s} + \frac{s}{2} \qquad (8)$$

here R is radius of substrate (circular) and s is the sag in the curvature profile. The surface profiles of substrate before and after deposition for GLAD film representatives, Samp-1 (0.5 rpm), Samp-2 (1 rpm), Samp-4 (3 rpm) and Samp-5 (4 rpm), are shown in Fig. 10 (a, b, c, d) respectively. Changes in surface profile after coatings are visible. The value of substrate parameters for the calculation of stress are: $E_s$=188 GPa[35], $\nu_s$=0.28, and $t_s$= 0.3mm for stress estimation of $ZrO_2$ films. ND $ZrO_2$ film depicts high compressive stress with value, - (5259 ± 500) MPa whereas GLAD films depict very low stress varying between -60 MPa to +184 MPa. As can be seen from FESEM images shown in Fig.7, ND film depicts a continuous and dense microstructure with no visible columns. High compressive stress can be explained in terms of either by atomic peening[36-38] or ad-atom diffusion.[36,39-41] Atomic peening governs the stresses where high energy incoming atom or particle strikes the growing surface, causing local atomic displacement and hence densification of films. Ad-atom diffusion model assume that excess ad-atoms get incorporated in grain boundaries and leads to the higher densities than expected in equilibrium. However, these two methods of film densification are different. Ad-atom diffusion occurs under high atomic mobility over surface (high temperature) while atomic peening dominates in case of high kinetic energy of ad-atom. So, there is always a competition between these two processes responsible for densification in thin films. Densification of thin films leads to large compressive stress due to reduction in inter-atomic distance (densification) and defects at grain boundary (due to grain boundary densification)[36,42]. In present case, we have used low substrate temperature sputtering (80ºC) and hence atomic peening process is mainly responsible for highly compressive ND $ZrO_2$ film. For GLAD film, reduction in stress could be attributed to the evolution of porous columnar microstructure caused by prominent atomic shadowing effects. In GLAD films, inter-atomic distance between the atoms on the edges of columns and facing each other becomes comparable or larger than the equilibrium inter-atomic distance and columns become in tension due to resulting inter-atomic attraction between the columns. The variation of the estimated stress in the $ZrO_2$ GLAD films is shown in Fig. 11. Samp-1 with the least substrate rotation speed depicts a compressive stress of -63 MPa and as the rotation speed further increased, stress switches from compressive to tensile with gradually increasing magnitude. This variation can be explained in terms of variation of inter-columnar distance in GLAD microstructure with substrate rotation. As described in section 3.1, with substrate rotation, columns get finer with the increase in substrate rotation and consequently average inter-columnar distance increases and becomes more than equilibrium position. So, such increase in inter-columnar distance with substrate rotation leads to increase in tensile stress. But, Samp-5 (4rpm) which depicts the lowest packing density depicts lower tensile stress than Samp-4 (3rpm). This can be attributed to the decrease in slope of inter-atomic attractive forces with high inter-atomic separation[42,43].

## 4. Conclusion

Deposition of $ZrO_2$ thin films in glancing angle and normal deposition configuration has been carried out on Si (111) substrate by RF magnetron sputtering technique at room temperature. The influence of substrate rotation on various properties of glancing angle deposited (GLAD) ZrO$_2$ films has been investigated. Various properties of GLAD films have been compared with that of ND ZrO$_2$ film. Optical properties viz., refractive index, extinction coefficient, and film thickness have been obtained through suitable modeling of spectroscopic ellipsometric measurements. GLAD films depict a substantial lowing in refractive index in comparison of ND films and the lowering is the attribute of dominant atomic shadowing which lead to porous columnar microstructure. Further, effect of substrate rotation speed on refractive index for GLAD films has been investigated and it follows a decreasing trend. Decrease in refractive index has been explained by exploiting the change in columnar microstructure in terms of decrease in column diameter. Deposition rate depicts an increasing trend with substrate rotation and it corroborates the decrease in column diameter with substrate rotation and hence decreasing trend of refractive index. Morphological parameters viz., correlation length, rms roughness and roughness exponent has been obtained through suitable modeling of height-height correlation function derived from atomic force microscopy measurements. Correlation length depicts a decreasing behavior with substrate rotation for GLAD films. Decreasing trend can be the attribute of decrease in columnar diameter with substrate rotation. AFM images also depict an evolution of surface grain from non-circular to circular shape as the substrate rotation increases. Trend of correlation length and AFM surface topography corroborate the variation of refractive index of the films. Grazing angle X-ray diffraction of ZrO$_2$ films depicts a preferential growth of GLAD films in monoclinic phase with (110) direction whereas ND film depicts a strong preferential growth in m (-111) direction. Preferential growth in different directions can be attributed to different angle of deposition which leads to the highest growth in perpendicular direction to the substrate for different planes for GLAD and ND ZrO$_2$ films. ZrO$_2$ films also depicts the presence of tetragonal peak with t (11-1) direction and fine nano-crystalline structure is responsible for the occurrence of room temperature tetragonal phase. Residual stresses in the films depict a switching from compressive to tensile stress. ND film possesses high compressive stress (5259 MPa) whereas stress for GLAD films lies between -60 to 184 MPa. High compressive stress of ND film is the attribute of atomic peening which leads to densification. Variation of stress for GLAD films has been explained from dependency of stress on inter-atomic distance in attractive region of inter-atomic forces. This study could play an important role in the development of challenging interference based multilayer devices and other technologies.

## References:

1. K. Robbie and M. J. Brett, Journal of Vacuum Science and Technology A **15,** 1460 (1997).
2. M. T. Taschuk, M. M. Hawkeye, and M. J. Brett, in *Handbook of Deposition Technologies for Films and Coatings (Third Edition)*, edited by P. M. Martin (William Andrew Publishing, Boston, 2010), p. 621.
3. A. Barranco, A. Borras, A. R. Gonzalez-Elipe, and A. Palmero, Progress in Materials Science **76,** 59 (2016).
4. M. M. Hawkeye, M. T. Taschuk, and M. J. Brett, in *Glancing Angle Deposition of Thin Films* (John Wiley & Sons, Ltd, 2014), p. 1.
5. R. B. Tokas, S. Jena, P. Sarkar, S. R. Polaki, S. Thakur, S. Basu, and N. K. Sahoo, Materials Research Express **2,** 035010 (2015).
6. H. WENDEL, H. HOLZSCHUH, H. SUHR, G. ERKER, S. DEHNICKE, and M. MENA, Modern Physics Letters B **04,** 1215 (1990).
7. S. Wang, G. Xia, X. Fu, H. He, J. Shao, and Z. Fan, Thin Solid Films **515,** 3352 (2007).
8. H. J. Qi, L. H. Huang, Z. S. Tang, C. F. Cheng, J. D. Shao, and Z. X. Fan, Thin Solid Films **444,** 146 (2003).
9. S. Shao, Z. Fan, J. Shao, and H. He, Thin Solid Films **445,** 59 (2003).
10. M. Levichkova, V. Mankov, N. Starbov, D. Karashanova, B. Mednikarov, and K. Starbova, Surface and Coatings Technology **141,** 70 (2001).
11. Y. J. Park, K. M. A. Sobahan, and C. K. Hwangbo, Optics Express **16,** 5186 (2008).
12. I. J. Hodgkinson, F. Horowitz, H. A. Macleod, M. Sikkens, and J. J. Wharton, Journal of the Optical Society of America A **2,** 1693 (1985).
13. R. M. A. Azzam and N. M. Bashara, *Ellipsometry and polarized light* (North-Holland Pub. Co., 1977).
14. S. M. Haque, K. D. Rao, J. S. Misal, R. B. Tokas, D. D. Shinde, J. V. Ramana, S. Rai, and N. K. Sahoo, Applied Surface Science **353,** 459 (2015).
15. H. Fujiwara, *Spectroscopic Ellipsometry: Principles and Applications* (Wiley, 2007).

| | |
|---|---|
| 16 | A. Garahan, L. Pilon, J. Yin, and I. Saxena, Journal of Applied Physics **101,** 014320 (2007). |
| 17 | A. S. Ferlauto, G. M. Ferreira, J. M. Pearce, C. R. Wronski, R. W. Collins, X. Deng, and G. Ganguly, Journal of Applied Physics **92,** 2424 (2002). |
| 18 | G. E. Jellison and F. A. Modine, Applied Physics Letters **69,** 371 (1996). |
| 19 | M. Senthilkumar, N. K. Sahoo, S. Thakur, and R. B. Tokas, Applied Surface Science **252,** 1608 (2005). |
| 20 | R. B. Tokas, S. Jena, S. Thakur, and N. K. Sahoo, Thin Solid Films **609,** 42 (2016). |
| 21 | B. Dick, M. J. Brett, and T. Smy, Journal of Vacuum Science & Technology B **21,** 2569 (2003). |
| 22 | Q. Zhong, D. Inniss, K. Kjoller, and V. B. Elings, Surface Science Letters **290,** L688 (1993). |
| 23 | J. Ferré-Borrull, A. Duparré, and E. Quesnel, Applied Optics **40,** 2190 (2001). |
| 24 | C. H. Zhang, Z.-J. Liu, K. Y. Li, Y. G. Shen, and J. B. Luo, Journal of Applied Physics **95,** 1460 (2004). |
| 25 | A. Mannelquist, N. Almqvist, and S. Fredriksson, Applied Physics A **66,** S891 (1998). |
| 26 | T. Karabacak, Journal of Nanophotonics **5,** 052501 (2011). |
| 27 | M. Jafarpour, E. Rezapour, M. Ghahramaninezhad, and A. Rezaeifard, New Journal of Chemistry **38,** 676 (2014). |
| 28 | J. Chevalier, L. Gremillard, A. V. Virkar, and D. R. Clarke, Journal of the American Ceramic Society **92,** 1901 (2009). |
| 29 | B. Alzyab, C. H. Perry, and R. P. Ingel, Journal of the American Ceramic Society **70,** 760 (1987). |
| 30 | Z. Ji and J. M. Rigsbee, Journal of the American Ceramic Society **84,** 2841 (2001). |
| 31 | R. C. Garvie, The Journal of Physical Chemistry **69,** 1238 (1965). |
| 32 | L. Zhang, H. Yang, X. Pang, K. Gao, and A. A. Volinsky, Surface and Coatings Technology **224,** 120 (2013). |
| 33 | D. Malacara, *Optical Shop Testing* (Wiley, 2007). |
| 34 | R. S. Sirohi, *Wave Optics And Its Applications* (Orient BlackSwan, 1993). |
| 35 | M. A. Hopcroft, W. D. Nix, and T. W. Kenny, Journal of Microelectromechanical Systems **19,** 229 (2010). |
| 36 | A. J. Detor, A. M. Hodge, E. Chason, Y. Wang, H. Xu, M. Conyers, A. Nikroo, and A. Hamza, Acta Materialia **57,** 2055 (2009). |
| 37 | N. A. Marks, D. R. McKenzie, and B. A. Pailthorpe, Physical Review B **53,** 4117 (1996). |
| 38 | C. A. Davis, Thin Solid Films **226,** 30 (1993). |
| 39 | C.-W. Pao, S. M. Foiles, E. B. Webb, D. J. Srolovitz, and J. A. Floro, Physical Review Letters **99,** 036102 (2007). |
| 40 | M. J. Buehler, A. Hartmaier, and H. Gao, Journal of the Mechanics and Physics of Solids **51,** 2105 (2003). |
| 41 | E. Chason, B. W. Sheldon, L. B. Freund, J. A. Floro, and S. J. Hearne, Physical Review Letters **88,** 156103 (2002). |
| 42 | A. Misra and M. Nastasi, Journal of Materials Research **14,** 4466 (1999). |
| 43 | A. Banerjea and J. R. Smith, Physical Review B **37,** 6632 (1988). |

**Caption of figures and tables:**

**Fig.1:** Photograph of glancing angle deposition configuration of RF magnetron sputtering system.

**Fig. 2 (a), 2(b):** Measured and fitted ellipsometric spectra of parameters, α (45º) and β (45º) for Samp-1(0.5 rpm substrate rotation) and Samp-6 (normally deposited). In inset, model layer structures are shown.

**Fig. 3 (a), 3(b):** Display dispersive curves of effective refractive index and its variation with substrate rotation for all the GLAD $ZrO_2$ thin films.

**Fig. 4(a), 4(b):** Fig. 4(a) presents measured height-height correlation function for all the $ZrO_2$ thin films including normally deposited. Fig. 4(b) shows measured and fitted height-height correlation function for Samp-1 (0.5 rpm).

**Fig. 5:** Plot of variation of correlation length for glancing angle deposited $ZrO_2$ thin films with substrate rotation.

**Fig. 6 (a, b, c, d, e, f):** 2-D AFM images of GLAD and normally deposited $ZrO_2$ thin film. Zoomed 3-D morphology has also been shown on the right in Fig. 6 to highlight grain geometry.

**Fig. 7:** FESEM cross-sectional images of glancing angle and normally deposited $ZrO_2$ thin films.

**Fig. 8:** Measured and simulated spectra of back scattered proton for normally deposited $ZrO_2$ film.

**Fig. 9:** Grazing incidence X-ray diffraction pattern of glancing angle and normally deposited thin films.

**Fig. 10 (a, b, c, d):** Surface contours of Si substrate before and after deposition of Samp-1 (0.5 rpm), Samp-2 (1.0 rpm), Samp-4 (3.0 rpm) and Samp-5 (4.0 rpm) respectively.

**Fig. 11:** Plot of residual stress vs. substrate rotation for glancing angle deposited $ZrO_2$ thin films.

**Table-1:** Optical parameters and film thickness derived from ellipsometric characterization.

**Table-2:** Morphological parameters and film thickness estimated from AFM and FESEM measurements.

**Fig. 1:**

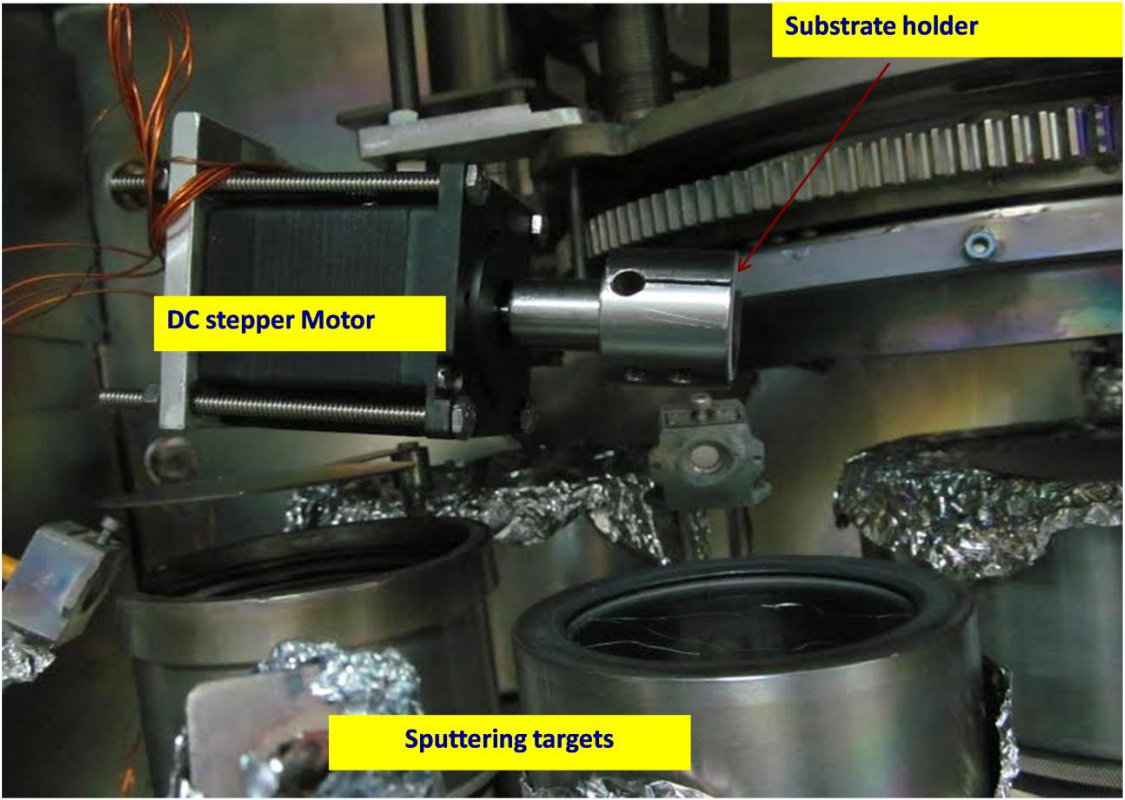

**Fig. 2(a):**

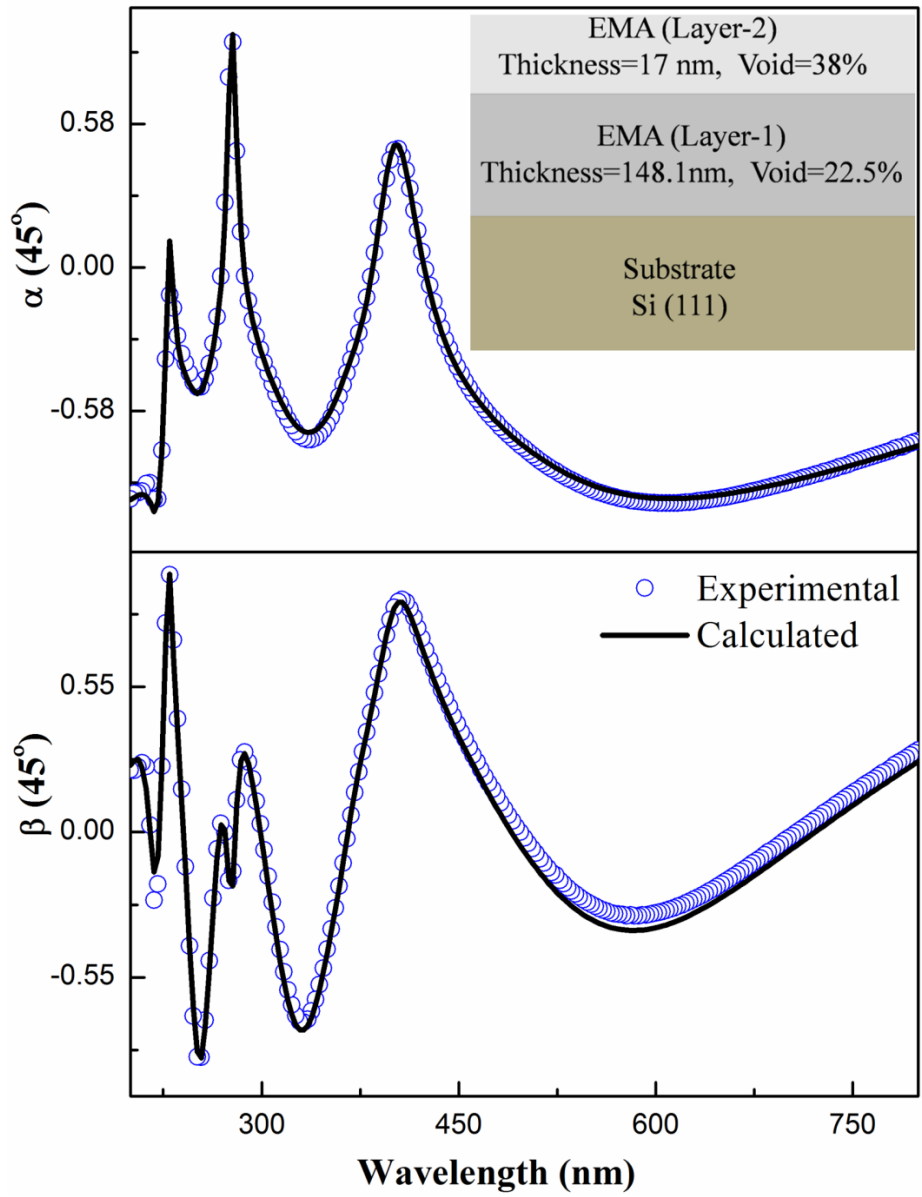

**Fig. 2(b)**

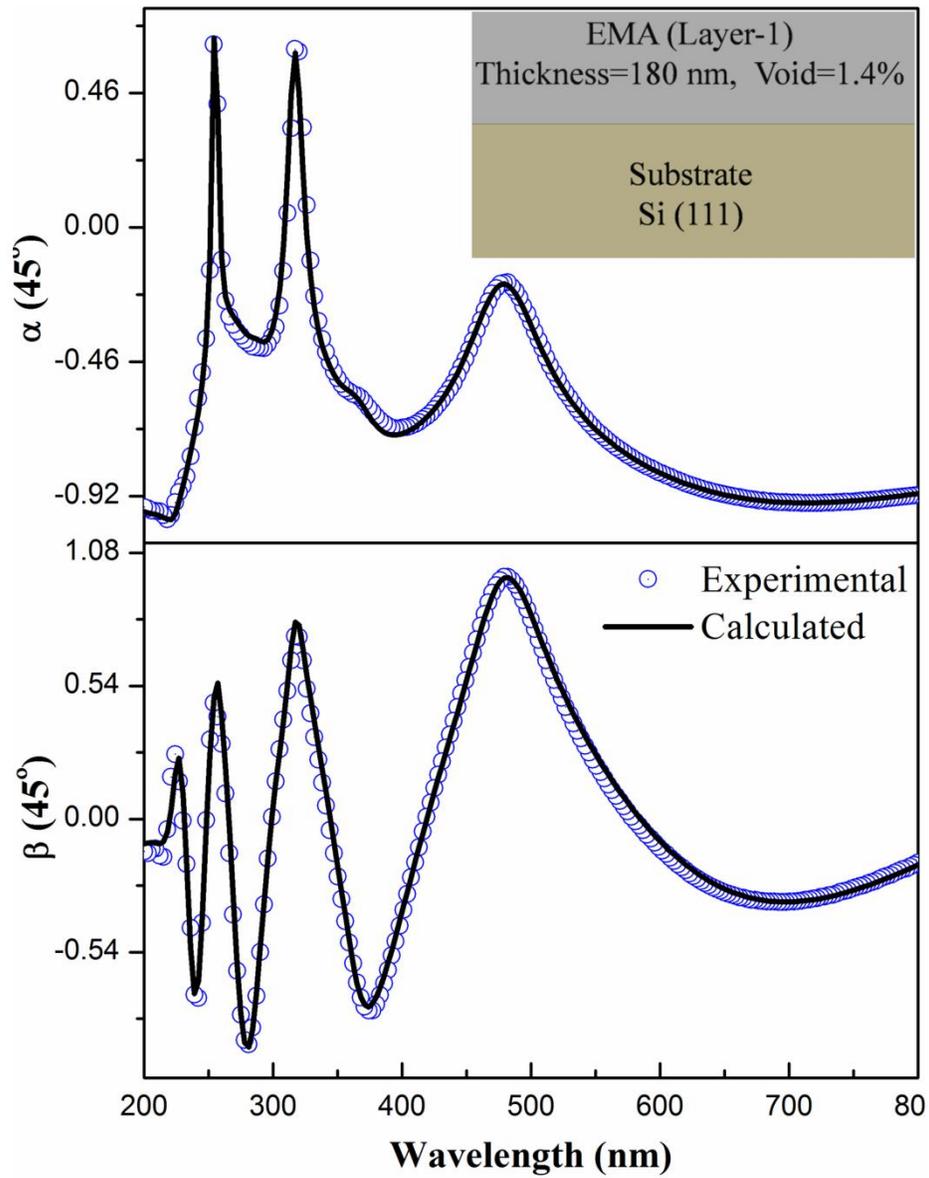

**Fig. 3(a):**

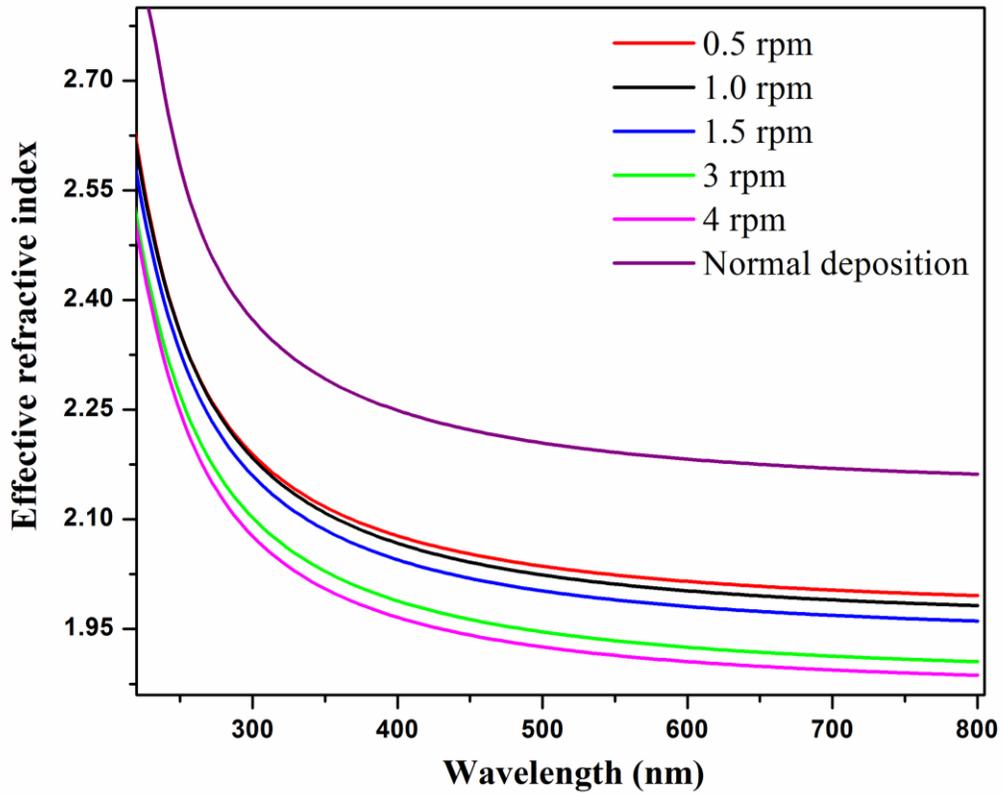

**Fig. 3(b):**

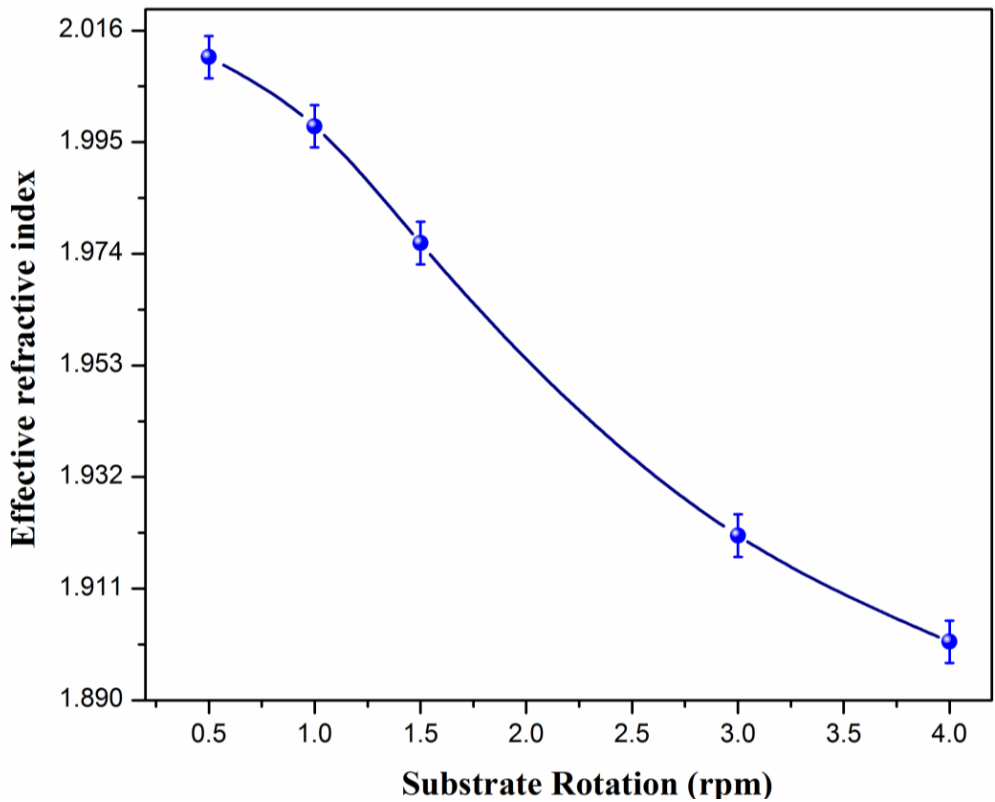

**Fig. 4(a)**

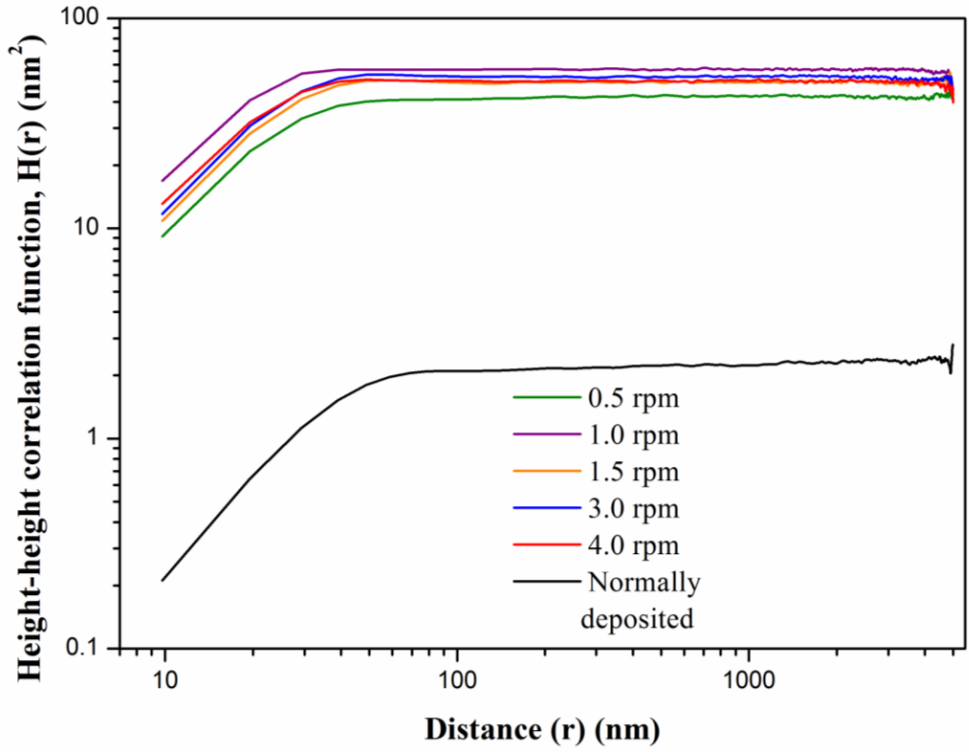

**Fig. 4(b):**

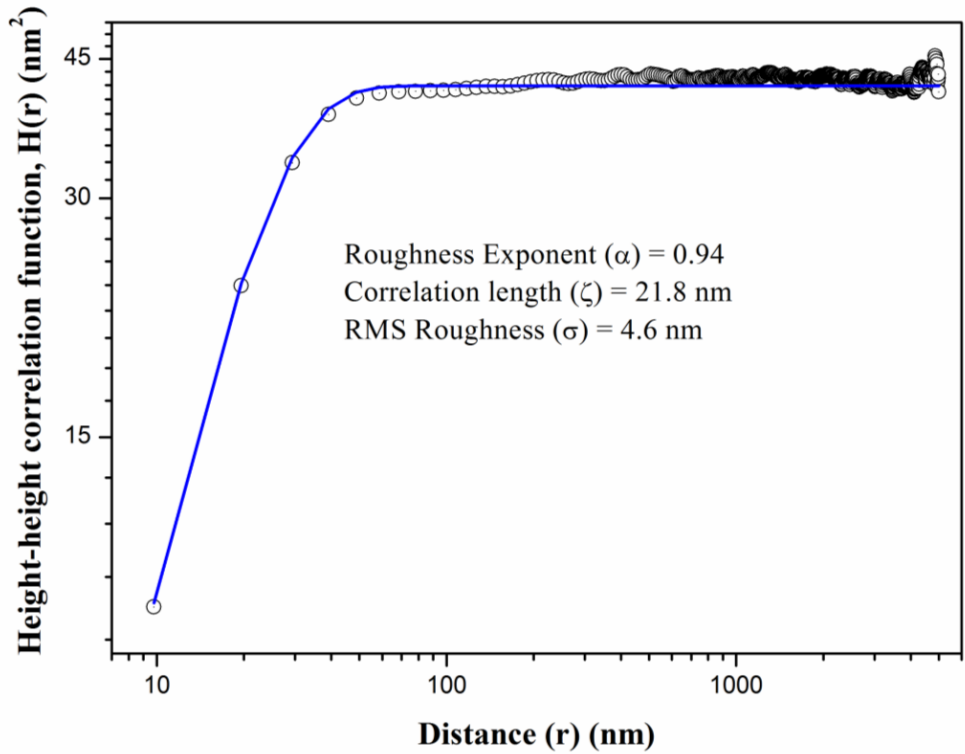

**Fig. 5:**

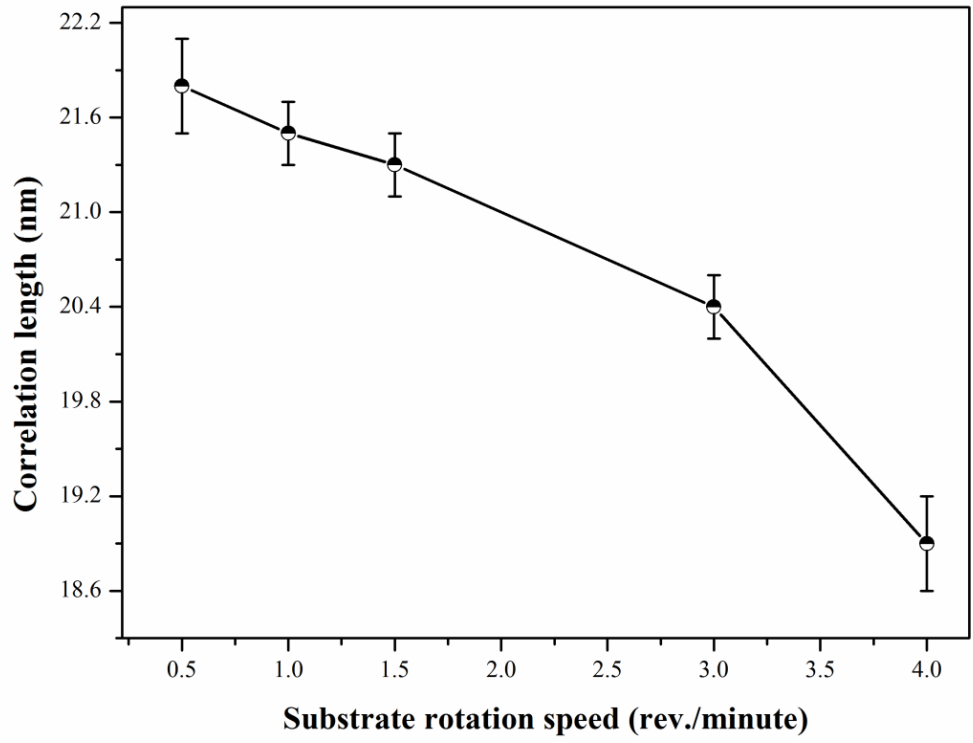

**Fig. 6**

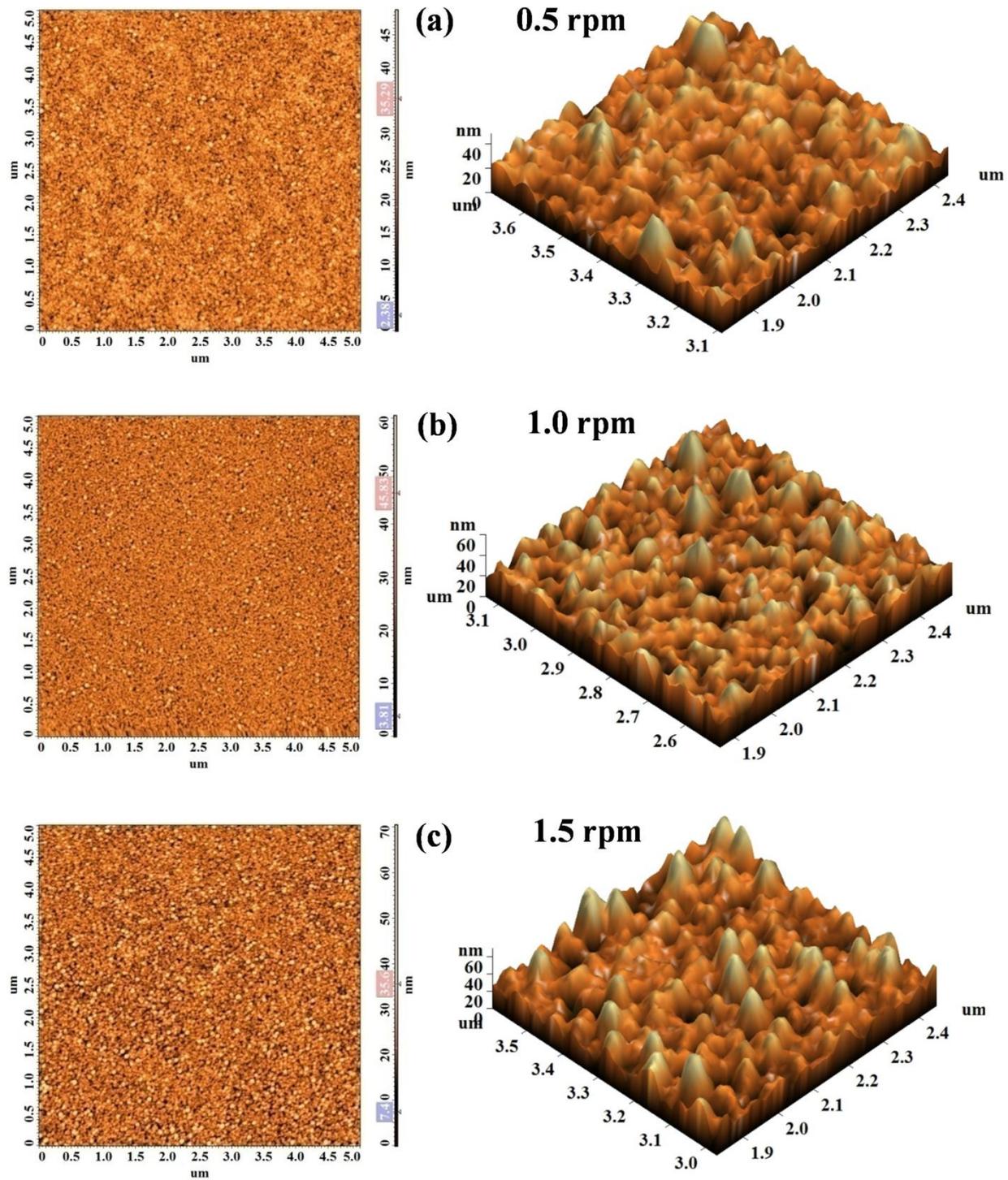

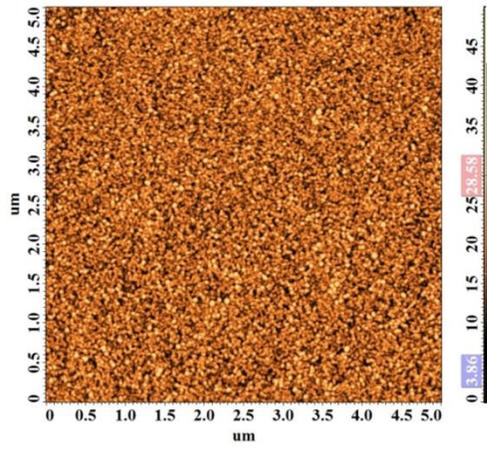 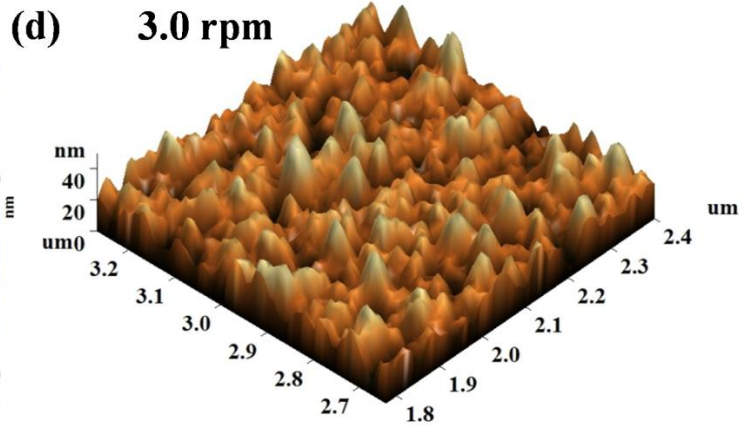

(d) 3.0 rpm

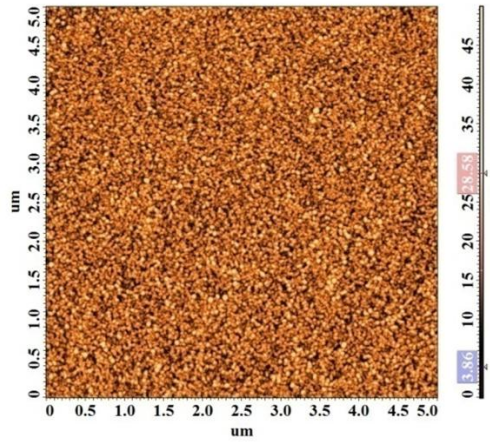 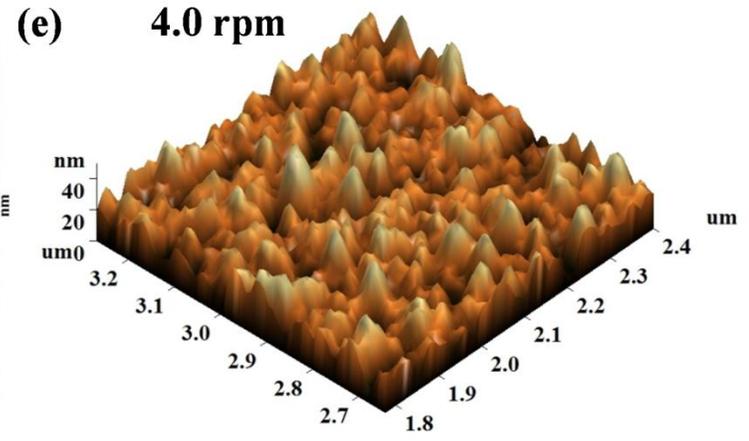

(e) 4.0 rpm

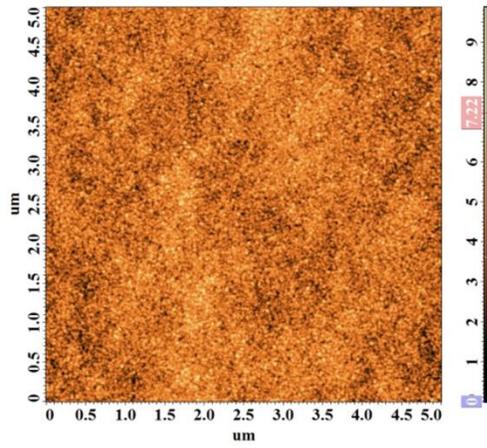 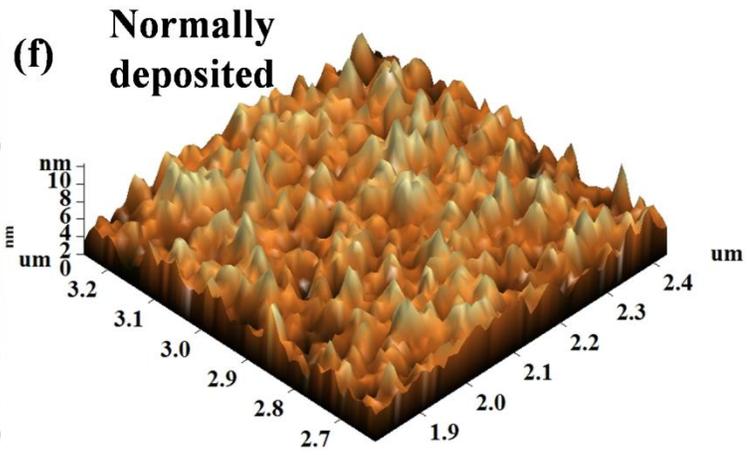

(f) Normally deposited

**Fig. 7:**

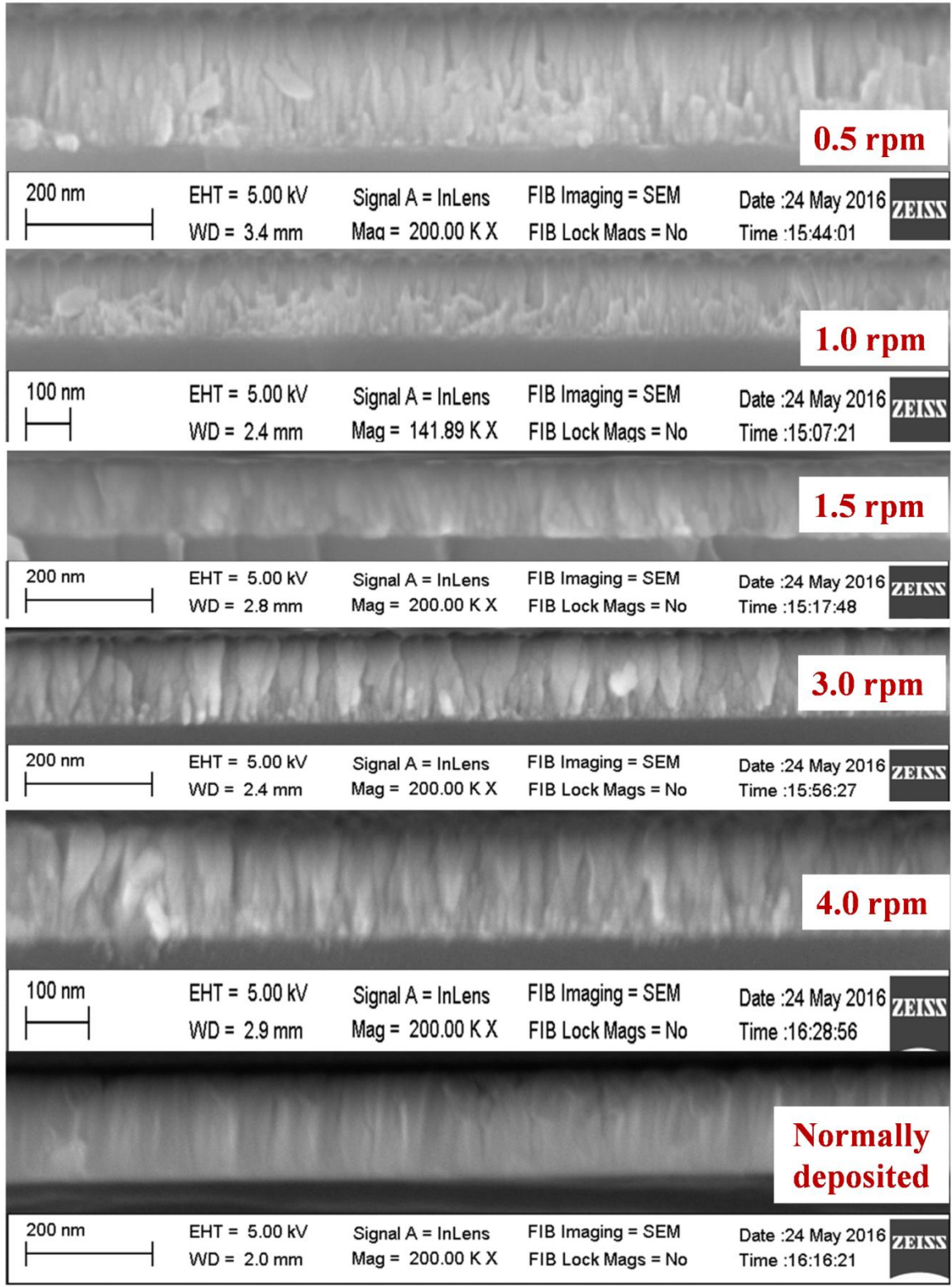

**Fig. 8**

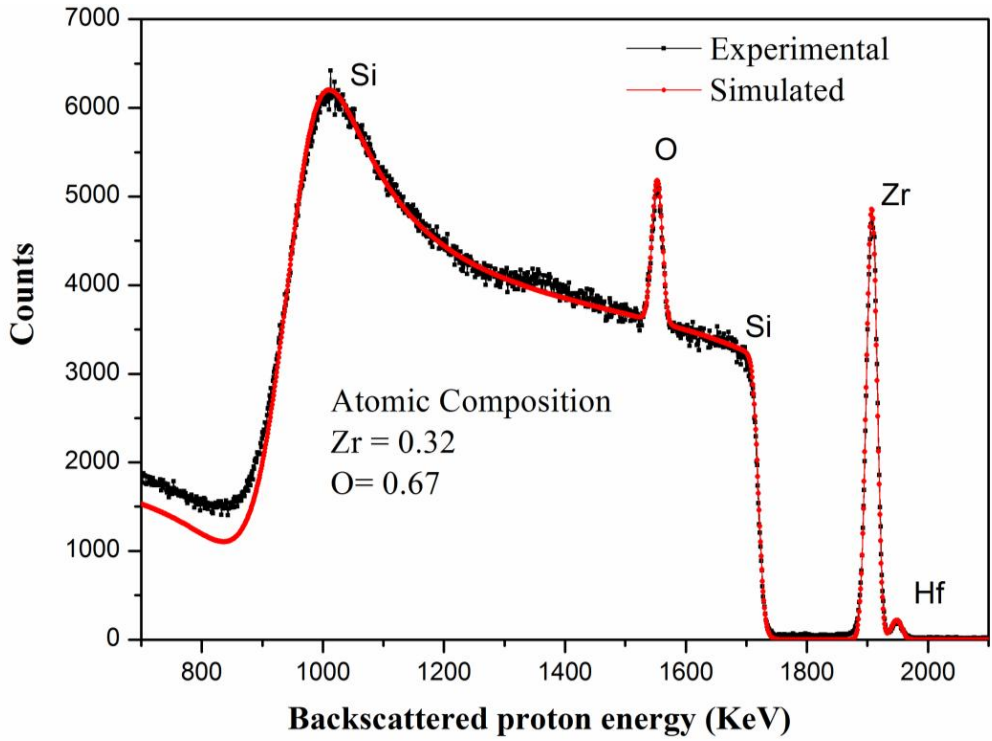

**Fig. 9**

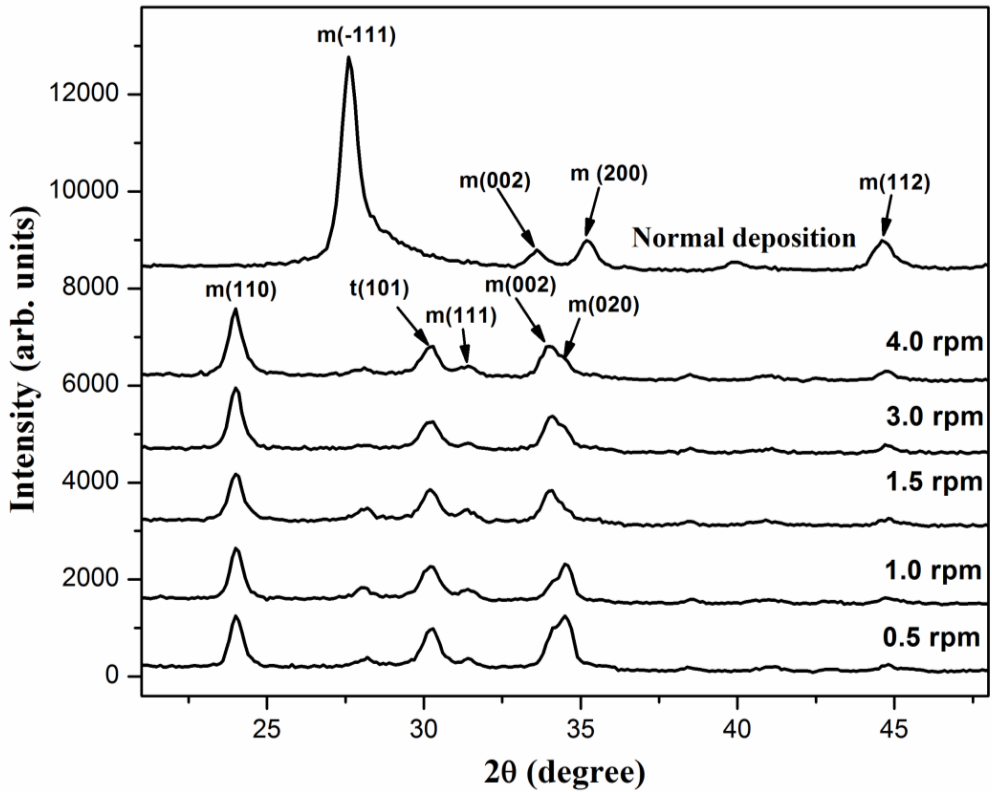

**Fig. 10:**

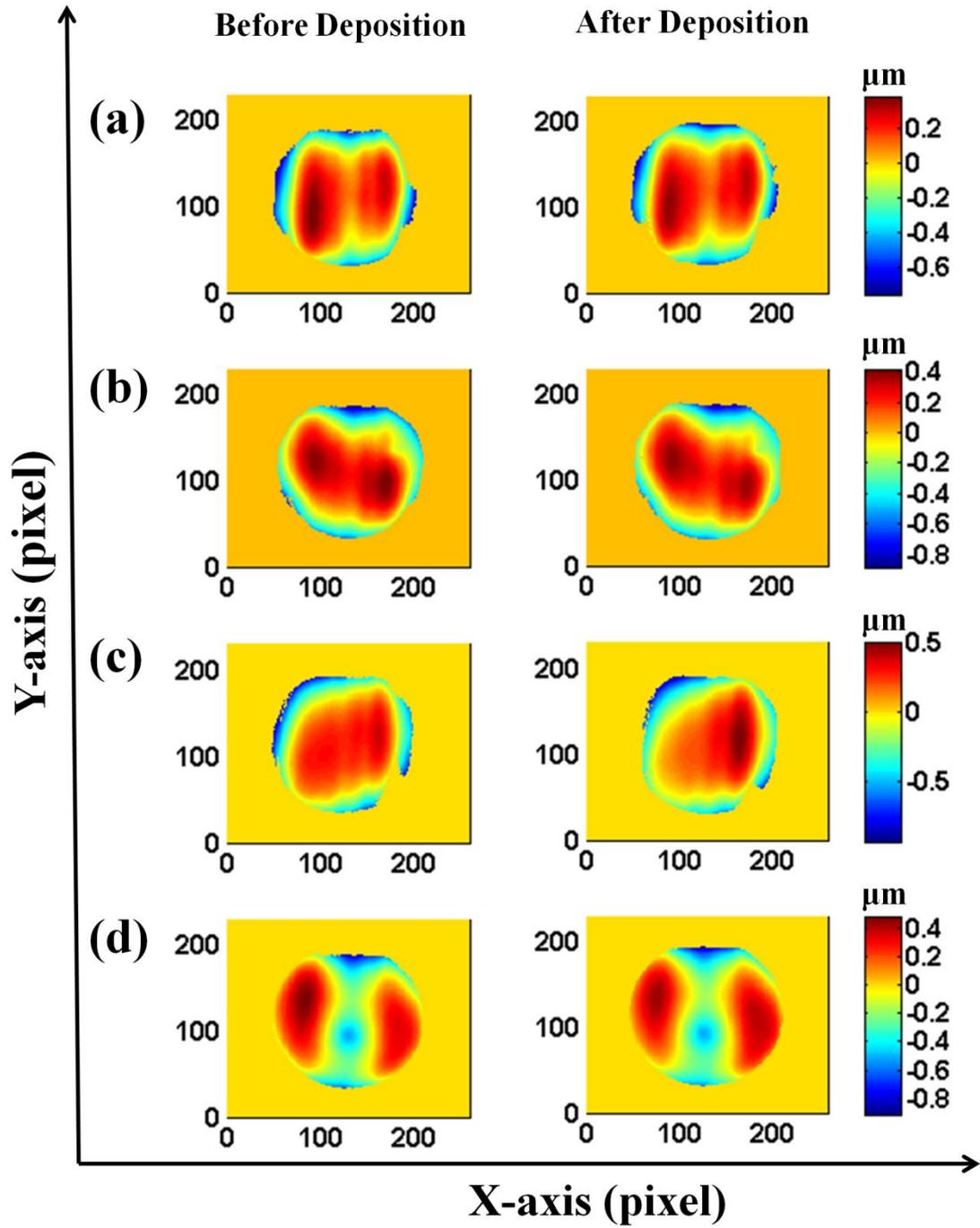

**Fig. 11:**

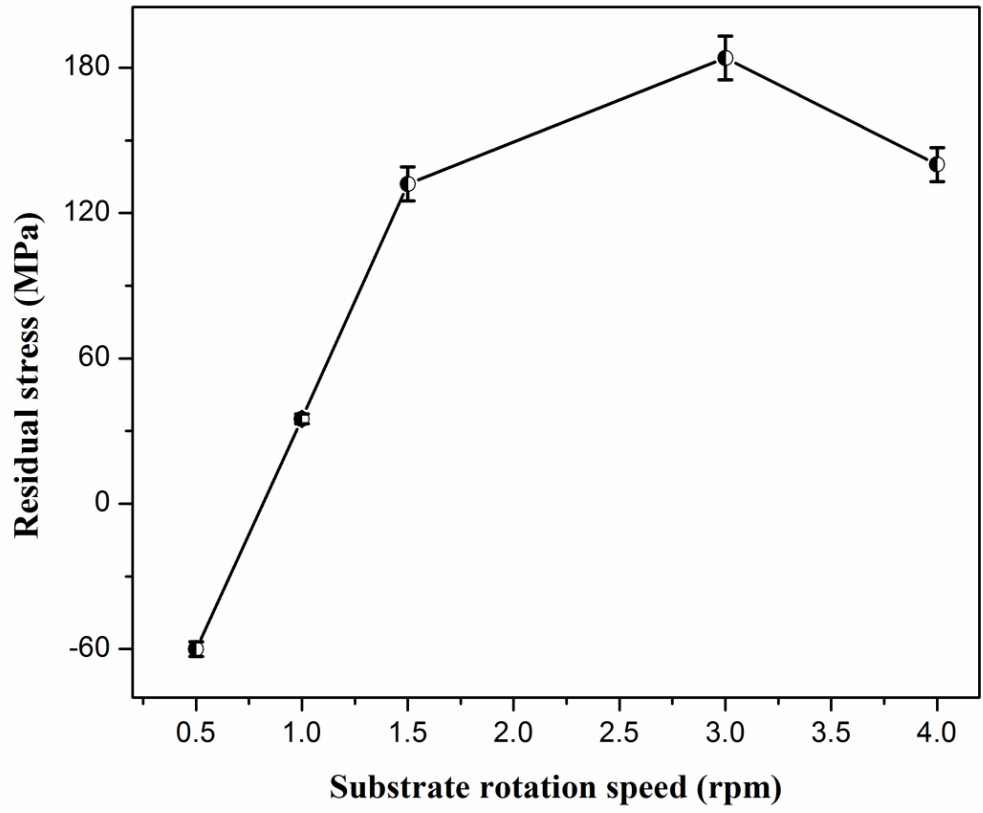